
\input phyzzx
\overfullrule 0pt
\titlepage
\Pubnum{SCIPP 91/16}
\date{June, 1991}
\pubtype{ T}     
\title{{Supersymmetry, Naturalness, and Dynamical Supersymmetry Breaking}
\foot{Work supported in part by the U.S. Department of Energy.}}
\author{Michael Dine
and Douglas A. MacIntire}
\address{}
%
\def\SCIPP{\centerline {\it Santa Cruz Institute for Particle Physics}
  \centerline{\it University of California, Santa Cruz, CA 95064}}
\SCIPP

\vskip.5cm
\vbox{
\centerline{\bf Abstract}
Models with dynamical supersymmetry breaking
have the potential to solve many of the naturalness problems
of hidden sector supergravity models.
We review the argument that
in a generic supergravity theory in which supersymmetry is
{\it dynamically} broken in  the hidden sector,
only tiny Majorana masses for gauginos
are generated.  This situation is
similar to that of theories with continuous R-symmetries,
for which Hall and Randall have suggested that
gluino masses could arise through mixings with an
octet of chiral fields.  We note that in
hidden sector models, such mixing can only occur if the auxiliary D field
of a $U(1)$ gauge field has an expectation value.  This
in turn gives rise to a catastrophically large Fayet-Iliopoulos
term for ordinary hypercharge.  To solve this problem
it is necessary to unify
hypercharge at least partially in a non-Abelian group.
We consider, also,
some general issues
in models with continuous or discrete R symmetries, noting
that it may be necessary
to include $SU(2)$
triplet fields, and that these are subject to various constraints.
In the course of these discussions, we consider a number of
naturalness problems.  We suggest
that the so-called ``$\mu$-problem" is not a problem, and point out
that in models in which the axion decay constant is directly related
to the SUSY breaking scale, squarks, sleptons and Higgs particles
generically acquire huge masses.}

\submit{Physical Review D.}
\endpage


\parskip 0pt
\parindent 25pt
\overfullrule=0pt
\baselineskip=18pt
\tolerance 3500

\endpage
\pagenumber=1

\chapter{\bf Introduction}
\bigskip

Hidden sector supergravity models provide a framework in which
to understand how various types of soft supersymmetry breaking
couplings might arise at low energies.  Unfortunately,
none of the models which have been
constructed to date
are at all compelling.  None are beautiful, and all suffer
from serious naturalness problems.
The most severe of these
is the problem of the cosmological constant.
Others include too large flavor changing neutral currents and neutron
electric dipole moment, the existence of a large hierarchy, put in by
hand,
and the need to omit from the lagrangian numerous couplings
permitted by symmetries, involving both visible sector and hidden
sector particles.

\REF\bt{M. Dine, A. Kagan and S. Samuel, Phys. Lett. {\bf 243B} (1990)
250.}
About the cosmological constant, we will have nothing new
to say here.  We will have to simply assume that this problem is
solved by some mechanism which does not too drastically alter
the low energy structure of the theory.
One might hope to
find conventional field-theoretic explanations for the other
questions.  If this is the case, these might have
phenomenologically interesting consequences, leading to predictions
concerning the spectrum of supersymmetric particles.
About flavor changing
currents and CP, for example,
it has been noted elsewhere\refmark{\bt} that
this problem might be resolved if the gauginos are the most
massive supersymmetric particles.  The problems of obtaining a large
hierarchy and of the omission of numerous couplings
in the hidden sector might be resolved if supersymmetry
\REF\dsbone{I. Affleck, M. Dine, N. Seiberg, Nucl. Phys. {\bf B241}
(1984) 493.}
\REF\dsbtwo{I.  Affleck, M. Dine, N. Seiberg, Phys. Rev. Lett.
{\bf 137B} (1984) 187.}
\REF\dsbthree{I. Affleck, M. Dine, N. Seiberg, Nucl. Phys. {\bf B256}
(1984) 557.}
is dynamically broken.\refmark{\dsbone-\dsbthree}
The need to omit various visible sector couplings,
on the other hand, might be resolved by the recent suggestion
of Hall and Randall that one impose an R symmetry on the
\REF\hr{L. Hall and L. Randall, Nucl. Phys. {\bf B352} (289) 1991.}
theory.\refmark{\hr}

Theories with dynamical supersymmetry
breaking (DSB) have been known for some time.\refmark{\dsbone-\dsbthree}
DSB
has the potential to explain large hierarchies,
and the known examples have the virtue that it is not necessary
to omit couplings allowed by symmetries.
However, if one proceeds in
the most straightforward way to build models based on these,
one runs into difficulties.  The most severe of these concerns
masses for gauginos, which turn out to be extremely small.
Of course, models with R symmetries also are in danger of
yielding small (zero) gaugino masses.  In the latter case,
Hall and Randall have proposed that the problem can be solved by
adding an octet of chiral fields to the low energy theory; these
particles combine with the gauginos to gain mass.  This mechanism
is a potential solution to the problems of DSB
as well.  We will see, however,
that in either case this mechanism can operate only
if the auxiliary D field
of a $U(1)$ gauge field in the hidden sector
acquires a large expectation value.  This in turn raises the
\REF\ps{J. Polchinski and L. Susskind, Phys. Rev.
{\bf D26} (1982) 3661.}
danger of a large Fayet-Iliopoulos term for hypercharge.\refmark{\ps}
One way (possibly the only way) to forbid such a term is to
unify hypercharge in a non-Abelian group at some scale.
We will see that the natural scale for breaking this additional
symmetry is the hidden sector scale.


While DSB may resolve some questions of naturalness in the hidden sector,
it is still usually necessary
to forbid
certain visible sector couplings.  For example, in the minimal
supersymmetric standard model, one must forbid a large mass term
in the superpotential for the two Higgs doublets, but one
must have a soft breaking mass term involving both doublets.
One can, as in the case of the unwanted hidden sector couplings,
simply suppose that the unwanted superpotential terms
are not present at tree level
and then invoke non-renormalization theorems.  Superstring theory
suggests that such a possibility might not be unreasonable.
Hall and Randall\refmark{\hr}
have recently considered an alternative
possibility, noting that continuous R symmetries can forbid such
terms.

\REF\apw{L. Alvarez-Gaume, J. Polchinski, and M. Wise,
Nucl. Phys. {\bf B221} (1983) 495.}
The rest of this paper is organized as follows.  In the
next section, we will briefly review some features of models
with dynamically broken supersymmetry.   In section 3, we will consider
what happens when such models are coupled to supergravity.
We will recall the general arguments that majorana
masses for gauginos in such theories must be small,
and illustrate them with a one loop calculation.  This
calculation is rather subtle; a conventional treatment,
such as has been applied to supergravity theories in the
past,\refmark{\apw}
gives a large mass.  It turns out, however, that the Feynman
diagrams contributing to the gaugino mass require careful
regularization, and that in the end these masses are small.
We comment on the implications of these results for more conventional
theories.

In section 4, we consider the effect of adding an
octet to a theory in which majorana masses for gauginos are small.
We first consider the problems associated with D terms, and possible
solutions.
We then consider models with dynamical
supersymmetry breaking
in which either gauge interactions or supergravity
is the ``messenger" of supersymmetry breaking.
It does not appear too difficult to build realistic models of this
type.

In section 5 we consider some aspects of models
in which majorana masses for gauginos are forbidden by
continuous or discrete R symmetries.  We point out, first, that
in supergravity theories, if one insists on cancelling the cosmological
constant in the effective lagrangian, only a discrete $Z_2$ R symmetry
can survive to low energies.  We argue, however, that given our
poor understanding of the cosmological constant problem,
and given all of the
naturalness problems of supersymmetric theories,
such symmetries are still worthy of study.
As noted by Hall and Randall, in such theories, in
addition to an octet of chiral fields, it may be necessary
to have still other fields to avoid very light states
in the neutralino sector.  These authors considered the
possible addition of a gauge singlet superfield.  We show
that such a singlet is unnatural, in the sense that
in almost any conceivable scheme for supersymmetry breaking,
it has unacceptable properties.  We note that the corresponding
problems do not arise for $SU(2)$ triplet fields, and
consider some aspects of such models, including the
spectrum and the question of the $\rho$ parameter.
We find that such schemes typically
predict that there should be new particles with
masses below $M_Z$.

\REF\kim{J. Kim, Phys. Lett. {\bf 136B} (1984) 378.}
\REF\wt{K. Rajagopal, F. Wilczek, W. Turner, Nucl. Phys. {\bf B358} (1991),
447.}
Our conclusions are presented in section 6.
Here we comment on possible connections of axion physics
and supersymmetry.  In particular, it is remarkable that
both of these require a scale of around $10^{11}$ GeV, and
a number of authors\refmark{\kim,\wt} have speculated on possible
connections between them.  We point out that generically, if
the Peccei-Quinn symmetry is broken by vev's in the hidden
sector, squark, slepton and Higgs masses tend to be of order
the intermediate scale.

\bigskip
\chapter{\bf Dynamical Supersymmetry Breaking}
\bigskip

\REF\wittenone{E. Witten, Nucl. Phys. {\bf B188} (1981) 513.}
\REF\nonrenorm{M.J. Grisaru, W. Siegel, and M. Rocek, Nucl. Phys.
{\bf 159} (1979) 429.}
\REF\wittentwo{E. Witten, Nucl. Phys. {\bf B202} (1982) 253.}
Witten was perhaps the first to appreciate the
possible importance of dynamical supersymmetry
breaking and to clearly formulate the problem.\refmark\wittenone\ He
stressed that
dynamical supersymmetry breaking was likely to give rise to large
hierarchies.  Because of the non-renormalization
theorems,\refmark\nonrenorm\ supersymmetry,
if unbroken at tree level, remains unbroken to any finite order
in perturbation theory.  However, he pointed out that
the proofs of the non-renormalization
theorems are firmly based on perturbation theory.
Thus one can hope to find effects smaller than any power
of the coupling constant which give rise to supersymmetry breaking.
Witten went on to prove that
many theories do not break
supersymmetry dynamically.\refmark\wittentwo\
However, chiral gauge theories did not yield to this analysis.

In a series of papers, it was in fact shown that supersymmetry
is sometimes dynamically broken in
four dimensions.\refmark{\dsbone-\dsbthree}  First it
was observed that non-perturbative breakdown of the non-renormalization
theorems is common -- almost generic.  The basic point is illustrated
by an $SU(2)$ gauge theory with a single massless flavor, i.e. containing
two chiral doublets, $Q$ and $\overline{Q}$.  At the classical level,
this theory has a continuum of physically inequivalent vacuum states.
Essentially these are the states with $Q = \overline{Q}=\left (
  \matrix{0 \cr v} \right )$.
For non-zero $v$, the gauge symmetry is completely broken, and the
gauge bosons are massive.  The effective coupling in a given vacuum is
$g(v)$,  since
the gauge boson masses are of
order $v$, and all momentum integrals are cut off in the
infrared at this scale.
As the
theory is asymptotically free, by choosing $v$ large enough the theory may be
made as weakly coupled as one wishes.
In each of these states there is one massless chiral field.  This
field can be
written in a gauge-invariant way as $\Phi=\overline{Q} Q$.  Expanding
the fields $Q$ and $\bar Q$ in small fluctuations about their vacuum
expectation values, the term linear in the fluctuations is the massless
state.
The problem, then, is to understand the properties of the effective
low energy theory containing $\Phi$ only, and in particular to determine
whether this theory possesses a superpotential for $\Phi$.
Symmetry considerations restrict the superpotential to be
of the form
$$W={\Lambda^5 \over \Phi}\eqn\superone$$
Here $\Lambda$ is the scale of this SU(2) theory, and
again, this expression should be
understood by expanding $\Phi$ in small fluctuations about a
particular ground state.It is straightforward to show that a single
instanton generates the various component interactions
implied by this superpotential.
This analysis immediately generalizes to
theories with gauge group $SU(N)$ and $N-1$ flavors.  Adding small
mass terms, one finds that all of these
results are consistent with Witten's analysis of dynamical
supersymmetry breaking.
Minimizing the full superpotential
yields $N$ gauge-inequivalent ground states, in agreement with
Witten's computation of the index.  By other methods, one can
show that a superpotential is generated in many other theories.

While these examples illustrate that the non-renormalization theorems
do indeed break down non-perturbatively, they do not lead to
a particularly interesting phenomenology.  Without mass terms, the potential
for
the field $\Phi$ falls rapidly to zero for large $\Phi$.  Thus the model
has at best a cosmological interpretation.
The basic problem is that
for large expectation values of the fields, the theories become
more weakly  coupled and any potential which is generated non-perturbatively
must tend to zero.
Adding mass terms to
the theory eliminates the ``flat directions" which exist classically
in the potential, but in this case the full theory has supersymmetric
ground states.  A general criterion for obtaining supersymmetry
breaking with a good ground state was suggested in ref. \dsbtwo.
Suppose a theory has, classically, no flat directions in its
potential.  At the same time suppose that the theory possesses a
continuous global symmetry which is broken in the true vacuum.  In such
circumstances supersymmetry is almost certainly broken.
For, if it were not, the goldstone bosons of the broken symmetry would have
scalar partners which would have no potential.  However, in this
case their expectation values would not be fixed and there would
be flat directions, contradicting the original assumption.

This argument is heuristic, and one can imagine a variety of loopholes.
However, a number of models were studied in ref. \dsbthree\
satisfying these
criteria, and shown to break supersymmetry.  The simplest is a theory
with gauge group $SU(3) \times SU(2)$, with chiral fields $Q$, $\bar U$,
$\bar D$
and $L$, transforming as
$(3,2)$, two $(\bar 3,1)$'s and $(1,2)$, respectively under the group.
In addition to the gauge interactions, to eliminate the flat directions
it is necessary to include a superpotential $$W= \lambda Q \bar Q L.
\eqn\supertwo$$  If $\lambda$ is small, one can first determine the
superpotential generated by instantons (as in equation \superone),
and then treat the tree level superpotential (equation \supertwo)
as a small correction.  Minimimizing the
resulting potential, one finds that supersymmetry is broken.
If the scale of the $SU(3)$, $\Lambda_3$, is larger than that of $SU(2)$,
$\Lambda_2$, one finds that at the minimum
$$Q=\left ( \matrix{a & 0 &  0 \cr 0 &  b & 0} \right ) ~~~~~~~~\bar Q
=\left ( \matrix{\bar U \cr \bar D}\right )= Q
 ~~~~~~~~~~~L=\left ( \matrix{\sqrt{a^2 -b^2} \cr 0}\right )
\eqn\formofminimum$$
where $a = 1.286{\Lambda_3 /\lambda^{1 \over 7}}$,
  $b=1.249{\Lambda_3 /\lambda^{1 \over 7}}$,
and the vacuum energy is $E=3.593 \lambda^{10 \over 7} \Lambda_3^4$.

Other models can be analyzed along similar lines.  Another example
of interest is an SU(5) theory with a single $\bar 5$ and $10$.
In this theory, there is no classical superpotential which one can
write.  Even so, the theory has no flat directions.  Using
't Hooft anomaly conditions one can argue that the
non-anomalous global symmetry of the model must be broken, and
that as a result supersymmetry is broken.

\bigskip
\chapter{\bf Coupling to Supergravity}
\bigskip

We would like to consider a theory of this type as a candidate
for the hidden sector of a supergravity model.\foot{As explained
in \dsbthree, breaking SUSY at low energies tends
to give unwanted axions and Goldstone bosons.}
As a concrete
example, we take the $SU(3) \times SU(2)$ model described in the
previous section; however, our considerations below generalize
almost trivially to other theories.
We assume
that, apart from some possible superheavy (${\cal O}(M_P)$ or
${\cal O}(M_{GUT})$) fields, no other fields transform under
the $SU(3) \times SU(2)$ gauge symmetry of the hidden sector
(these groups should {\it not} be confused with the $SU(3)
\times SU(2) \times U(1)$ symmetry in the visible sector; they
represent {\it additional} gauge interactions).  Thus, taking
the characteristic scale of the hidden sector, $M_{int}$, to be
$M_{int} \sim 10^{11} GeV$, the dynamics
described in the previous section are unaffected:
supersymmetry is broken in this sector at a scale of order $M_{int}$;
various fields acquire expectation values and masses of order $M_{int}$,
and there are some (pseudo) Goldstone fields with decay constants
of this order.

\FIG\gaugino{One loop diagram contributing to gaugino masses.}

The analysis of scalar masses is similar to the case of more
conventional hidden sector supergravity models.  The problem comes
when one attempts to compute gaugino masses.  There is a simple argument that
any Majorana masses for gauginos
in such theories must be extremely small.  In discussing physics
at scales above $m_{3/2}$,
it should be possible
to integrate out Planck (and GUT) scale physics, obtaining
a (locally) supersymmetric effective lagrangian.
The usual supergravity lagrangian is specified by three functions.
Only the function $f(\phi_i)$, which describes the coupling of the
chiral fields to the gauge multiplets, is relevant to the question
of gaugino mass through a term in the lagrangian:
$${\cal L}\sim \int d^2\theta f(\phi) W^\alpha W^\alpha$$
where $f$ is a holomorphic function of the scalar fields.
On the other hand,
  in all of the models of dynamical supersymmetry breaking
presently known, all of the hidden sector fields, $Z_i$, carry charges under
the various gauge symmetries.  Thus $f$ is necessarily at least
quadratic (and in fact is generally cubic) in fields.  Thus
one expects its coefficient to be suppressed by at least two
powers of $M_P$.  If this is the case, local supersymmetry implies that
gaugino masses will be extremely small (of order $eV$ or smaller).

However, if one simply computes the gaugino masses in these models
using the naive Feynman rules,
one seems to find much larger answers.   For example,
suppose that, in addition to a hidden sector of the type we have
described in the previous section,
the model possesses a heavy color octet, $O$, of chiral
fields of mass $M$.  Then at one loop there is a diagram
contributing to the gluino mass, quite similar to the types
of diagrams considered in ref. \apw.  In particular, there
is a (non-vanishing)
term in the lagrangian of the form $m_{3/2} M  O^2$,
where $O$ is the scalar component of the octet.
Then the diagram of fig. \gaugino\
gives a non-zero mass for the gluino of order
$$m_{\lambda} \sim {\alpha_3 \over \pi}
 {M_{int}^2 \over M_P}\eqn\gluino$$
Notice, in particular, that this expression is independent of the mass of
the octet.  If correct, this would be a wonderful result,
since it would give rise to a gluino mass of order $100 GeV$
of so.  Not surprisingly, in view of
our general argument, this result is not supersymmetric.

The problem with this calculation is most easily illustrated
with a
well-studied model, the ``Polonyi model."
This theory contains
a hidden sector consisting of only one singlet chiral
field, $Z$, with superpotential
$$W= M_{int}^2(Z+\beta) \eqn\polonyiw$$
Assuming that the Kahler potential is simply quadratic in $Z$,
the minimum of the potential occurs for
$$Z=(\sqrt{3}-1) M\eqn\zmin$$
where $M={M_p \over \sqrt{8 \pi}} $; in these equations,
in order that the cosmological constant vanish at the minimum
of the potential, $\beta= (2-\sqrt{3})M$.
Because $Z$ is a gauge singlet, there should be
no problem obtaining a gaugino mass in this model, since
any $f$ which is, say, a polynomial in $Z$ will yield $m_{\lambda}
 \sim m_{3 \over 2}$.
Indeed, if one now adds to this
model the heavy octet, $O$, above,
one generates a gluino mass at one loop;\refmark{\apw}
proceeding as before, the diagram of fig. \gaugino\ yields
$$m_{\lambda} ={3g^2 \over 4\pi^2}(2- \sqrt{3})
e^{(\sqrt{3}-1)^2 \over 2}
m_{3 \over 2} \eqn\polonyigluino$$
with $m_{3/2}\sim{M^2_{int}\over M_p}$.  Again we find a mass of the order of
100 GeV.

Now we would expect that at energy scales below the mass of the heavy
octet, $M_O$, the system would still be described by a locally
supersymmetric effective lagrangian, including the usual light
fields and the hidden sector fields.  In particular, the
gluino mass could be understood as arising from the
function $f$ of this theory, through the term in the supergravity lagrangian:
$${\cal L}\sim{1\over 4}e^{-G/2}G^l(G^{-1})^k_l f^*_{\alpha \beta
k}\lambda^\alpha \lambda^\beta.$$  In the above, $f^*_{\alpha \beta k}$ is the
derivative of $f^*_{\alpha \beta}$ with respect to $Z$, and the existence
of a gluino mass implies that $f$ is a function of $Z$.
If this is the case, on the other hand, we expect to find couplings
of $Z$ to $F_{\mu \nu}^2$ and $F \tilde F$ through the terms:
$${\cal L}\sim -{1\over 4} Re f_{\alpha\beta}F^\alpha_{\mu\nu}F^{\mu\nu\beta} +
{1\over 4} i Im f_{\alpha\beta}F^\alpha_{\mu\nu}\tilde F^{\mu\nu\beta}.$$
However, at one loop,
\REF\cremmer{E. Cremmer, B. Julia, J. Scherk, S. Ferrara,
L. Girardello and P. van Niewenhuizen, Nucl. Phys. {\bf B147} (1979)
105;
J. Wess and J. Bagger, {\it
Supersymmetry and Supergravity}, Princeton University Press,
Princeton (1981) and Johns Hopkins preprint (1990).}
using the lagrangian of refs. \cremmer,
these couplings vanish!

To see this, consider the coupling of
the pseudoscalar part of $Z$ to the octet fermions.
This coupling is proportional to $\partial_{\mu} Z
\bar O \gamma^{\mu} (1 - \gamma_5) O$.  One can attempt to
compute the coupling of the imaginary part of $Z$ to
$F \tilde F$ at one loop arising from this term.   But this
calculation is identical to the famous calculation of the
chiral anomaly, and is subject to the same ambiguities.
For example, it is well known that if one uses, say, a  Pauli-Villars
regulator, the $F \tilde F$ coupling vanishes in this case.
Indeed, the result of this computation, as in the case of the
gaugino mass above, is independent of the mass of the particle
running in the loop, and so is canceled by the regulator diagram.
Clearly supersymmetry requires that one use the same sort of regulator
for all of the diagrams.  But we have seen that the gluino mass
is independent of the mass of the heavy particle running in the loop,
so adding the Pauli-Villars term will give zero!  Correspondingly, alternative
choices
of (supersymmetric) regulators will give different results for
the gluino mass.  However, in the case of the hidden sector with
DSB, it is clear from our original symmetry arguments that any gauge-
and supersymmetric regulator will give zero for the gaugino mass.
\foot{We are assuming, here, that the lagrangian given in ref.
\cremmer\
is the most general one consistent with local supersymmetry, up to
terms with two derivatives.}

Thus simply using a theory with dynamical supersymmetry breaking
as a conventional hidden sector model yields unacceptable results.
In the following section, we consider an alternative approach.

\bigskip
\chapter{Models with Octets}
\bigskip

Majorana masses for gauginos are also forbidden in theories
with an exact R symmetry at low energies.  Following Hall and Randall,
it is natural to attempt
to build models with light color octet chiral fields, and to
allow them to mix with the gluinos.  In this section, we will consider
some general issues in models of this kind (with either DSB and/or
exact R symmetries).

As stressed by these authors,
a mass term mixing the gluino and the octet fermions
is one of the allowed soft breaking terms
of supersymmetry.  It is interesting to ask, on the other hand,
how such a term might arise in the framework of hidden
sector models.  Consider, first, the case of hidden sector models
with global supersymmetry.  (It is convenient to consider this
case because it is easy to write down globally supersymmetric
effective actions).  In such theories,
above the scale of weak interactions (the scale of supersymmetry
breaking in the visible sector of the theory), it is possible to describe
the theory by a supersymmetric effective action.\refmark{\ps}
Then supersymmetry breaking is the statement that, below the breaking
scale, the auxiliary ($F$)
component of some chiral superfield(s), $Z$, is non-vanishing, as
well, possibly, as the auxiliary ($D$) components of some gauge
fields.  In such theories, masses for the scalar components of
observable fields (denoted by $\phi$), arise through operators
of the type $\int d^4 \theta Z^{\dagger} Z \phi^{\dagger} \phi$;
Replacing $Z$ ($F_Z$)
by its expectation value immediately yields scalar masses.
On the other hand, terms which mix the fermionic components of the
octet, $O$, with the gluinos can only arise provided the theory contains
a $U(1)$ gauge field, $\tilde V$, whose auxiliary component,
$\tilde D$, has an
expectation value.
\foot{
This does not occur in the $SU(3) \times SU(2)$
model discussed earlier.  There we can gauge a $U(1)$.
However, it is
is necessary to include an additional
field to cancel anomalies.  It turns out that the sign of the charge
of this
field is such that its expectation value gives a vanishing expectation
value for $\tilde D$.  We know of no reason for this to be
true in general, however.}
Then the desired mixing can arise through the
operator
$${\cal L}_{\lambda} ={1 \over M}\int d^2 \theta \tilde W_{\alpha}
W_{\alpha}^a O^a\eqn\gluinooperator$$

Of course, this $U(1)$ cannot be ordinary hypercharge.  But the
large vev of $\tilde D$
raises the specter of a large Fayet-Iliopoulos term for
$D^Y$.
The dimension four operator $\int d^2 \theta
W^{Y}_{\alpha} \tilde W^{\alpha}$ gives a Fayet-Iliopoulos
term of order $\vev{\tilde D}$.  Such a coupling implies a large
negative mass-squared for scalars carrying hypercharge (of
order $M_{int}^2$), and potentially leads to an enormous
breaking of hypercharge.  One possible way to avoid this problem
is to unify hypercharge into a non-Abelian group, broken only at
some scale well below $M_p$.  For example, many authors,
motivated by string theory, have considered the possibility
that down to some scale there is an unbroken $SU(3)_c \times
SU(3)_L \times SU(3)_R$ symmetry.  In such a case, the Fayet-Iliopoulos
term could be highly suppressed.    In fact, one can even avoid the
problem if hypercharge is a sum of a $U(1)$ generator and a
non-Abelian generator.   In such a case, it can be natural for
some scalar field to gain a large vev, breaking some of the gauge symmetry
and leaving ordinary hypercharge.
\foot{As an example, one can consider a set of fields with the
quantum numbers of a $27$ of $E_6$, and suppose that the unbroken
group is $SU(2)_L \times SU(2)_R \times U(1) \times U(1)$.
Suppose that, apart
from the Fayet-Iliopoulos term for the $U(1)$, all fields have positive
soft-breaking mass terms, of order $m_{3/2}^2$,
except for the two $SU(3)\times SU(2)
\times U(1)$ singlets, which have negative mass-squared terms.
Then it is easy to check that there is a local minimum of the
potential at which the surviving gauge symmetry is $SU(3) \times
SU(2) \times U(1)$.}

These considerations can be immediately extended to the case
of local supersymmetry.  If one examines the lagrangian of ref.
\cremmer, one can see that there is only one term which gives rise to
a Dirac mass term mixing gauginos and matter fields, and this
is only non-vanishing if there is an expectation value for $\tilde D$.

In theories with dynamical supersymmetry breaking, having
obtained a sufficiently large gluino mass, we have
more or less phenomenologically acceptable models.
One still must check the neutralino sector.  If the
superpotential contains a term $m H \bar H$, with
$m \sim m_{3/2}$, the only light neutralino is the
photino.  We will argue later that a term of this size
will arise automatically in many circumstances.
The photino may gain a small mass from loops of light fields,
but it may be necessary to add additional light fields in order
to obtain an acceptable phenomenology.

\bigskip
\chapter{Models with Continuous R symmetries}
\bigskip

Our remarks in the previous section apply to models with dynamical
supersymmetry breaking and to theories with unbroken continuous
R symmetries.  In both types of models, the desired mixing arises
if an auxiliary D field has a non-zero vev, and one must insist
on at least a partial unification of hypercharge in a non-Abelian
group to avoid Fayet-Iliopoulos terms.  In this section, we consider
some further issues associated with R symmetries.
Such theories have
previously been carefully considered by Hall and Randall.\refmark\hr\
These authors
assumed that the symmetry was continuous.  The
Higgs fields were assigned $R=0$, while quark and lepton
superfields were assigned $R=1$.  In order to obtain a
gluino mass, they required that their models contain a color octet
chiral field with
$R=0$; they then noted that $\lambda^a \psi_O^a$, where $\psi_O$
is the fermionic component of $O$, is one of the allowed soft
breaking terms.
Hall and Randall also observed that if one does not add
additional fields, the model possesses, at tree level, a massless
photino and a massless higgsino.
At tree level, one can suppress the coupling of the massless
Higgsino to the Z, however one predicts too light a Higgs.
As a result, these authors considered theories with an additional
singlet field.  We will argue shortly
that in almost any scenario for supersymmetry breaking, this
is likely to lead to difficulties; instead one needs to add
$SU(2)$ triplet fields.  Hall and Randall have recently pointed
out that
once one loop corrections are accounted for, it may not be
necessary to include additional fields at all.  The point
is that the large radiative corrections to the Higgs mass
\REF\haber{H. Haber and R. Hempfling, Phys. Rev. Lett. {\bf 66}
(1991) 1815; J. Ellis, G. Ridolfi and F. Zwirner, Phys. Lett. {\bf
B257} (1991) 83.}
due to top quark loops which have been discovered recently\refmark\haber\
can avoid the light Higgs
\REF\hrprivate{L. Hall and L. Randall, private communication.}
problem, provided the top quark is heavy enough.\refmark\hrprivate\
Of course, dynamical supersymmetry breaking could operate in the
framework of such models as well.

 We would like to explore some aspects
of models of this type.  First, there are some questions
of ``philosophy" and naturalness which must be addressed.
For most particle theorists, continuous global symmetries
are anathema, and this might be viewed as an objection
to the work of ref. \hr.  However, in order to implement
the program of these authors, it is not necessary that
the R symmetry be continuous; it can in fact be discrete.
Discrete $R$ symmetries have a different status.  For example,
they arise frequently in string theory, where they
are usually (possibly always) discrete gauge symmetries.
For suitable $Z_N$, a discrete $Z_N$ R symmetry has consequences
very similar to that of a continuous R symmetry.

For both discrete or continuous R symmetries, however,
there is a puzzle.
Supersymmetry,
if it exists, is a local symmetry.  Thus the underlying theory
must be a supergravity theory.  In an $N=1$ supergravity theory,
supersymmetry breaking with vanishing energy at the minimum
of the potential requires
that the superpotential have a non-zero expectation value.
But such an expectation value necessarily breaks any R-invariance
(apart from $Z_2$ symmetries).
In simple models, this breaking of R invariance tends to be large,
and, for example, large Majorana mass terms for gauginos are generated.
\foot{The theories with dynamical supersymmetry breaking often have an
approximate R invariance in the low energy theory, even after
cancelling the cosmological constant,
but this does
not help with the basic problems of naturalness.}
Still, given our lack of understanding of the cosmological constant
problem,
the possibility of an unbroken R-invariance seems
worthy of investigation.

On the other hand, we would like to reconsider the motivations
for considering R-symmetric theories given in ref. \hr.
These authors argue that such symmetries would improve
the ``naturalness" properties of supersymmetric theories.
For example, they would forbid a term in the superpotential
of the form $\mu H_1 H_2$, where $H_i$ denote the two
Higgs doublets.  This argument is not particularly compelling.
{}From string theory, for example, we know that it is plausible
to have massless Higgs doublets at tree level and to
any finite order in perturbation theory.  The question,
then, is how large is $\mu$ once one takes into account
supersymmetry breaking.  The situation is most easily
described in global supersymmetry.  There, if the hidden
sector contains some fields, $Z_i$, with non-vanishing
$F$-components, $F\sim  M_{int}^2 \sim
m_W M$, then $\mu$ is generated by operators
of the form
$${\cal L}_{\mu} = {1 \over M}
\int d^4 \theta Z^{\dagger} H_1 H_2\eqn\muterm$$
Replacing the chiral field $Z$ by its vacuum expectation
value $Z = \dots \theta^2 \vev{F}$ gives
$\mu= {<F> \over M} \sim m_W$.
A number of authors have noted that these couplings can arise
in loops.  In supergravity theories, they generically
arise at tree level.  For example, in an $SU(5)$ theory
in which a $24$ couples to Higgs in the superpotential,
in such a way that the Higgs mass vanishes as $m_{3/2}
\rightarrow 0$, supersymmetry breaking shifts the $24$
vev by an amount of order $m_{3/2}$, giving rise to
$\mu \sim m_{3/2}$.  Thus, in a generic theory, the
``$\mu$-problem" does not appear to be a problem.

The question of motivation aside, models with R symmetry are quite
interesting.
Singlets, however, are likely to lead to difficulty in this context.
The problem is that the dimension four term in the effective
lagrangian,
$${\cal L}_{S} = \int d^4 \theta Z^{\dagger} S \eqn\singletmu$$
gives rise, effectively, to a superpotential term
$$W_{S}= \vev{F} S\eqn\singletw$$
Because $F$ is generically so large, this term generally has
disastrous consequences; for example, it leads to expectation
values for Higgs doublets of order the intermediate scale.
If one has, instead of singlets, some additional triplet fields,
this problem does not arise.  The corresponding ``$\mu$" term, as
for Higgs fields, is of order $m_W$.

Actually, in models with dynamical supersymmetry breaking, in contrast
to the more general case, this problem may be somewhat ameliorated.
The point is, again, that the $Z_i$'s are all charged under the
hidden sector gauge symmetries, so it is necessary to go to higher
dimension operators in order to find these $\mu$ terms.  Explicit
checks show that at one loop, such terms are indeed generated
only with suitably small coefficients.  Thus in this
framework, models with singlets may make sense.  However, it is of some
interest to explore the case of models with triplets as well.
This is rather straightforward extension of the work of ref. \hr,
which we now describe.

\def\L{{\cal{L}}}

In the case of triplets, there are a number of phenomenological concerns.
One has to insure that the triplet expectation values are small
enough that the $\rho$ parameter is not significantly affected.
Also, one must make sure that there are no particles so light
that their effects would already have been observed at LEP.
For definiteness, we will focus on the case where the
$R$ symmetry is continuous.

As usual, we assign all ordinary matter fields R-charge
zero; in other words, the chiral fields associated
with the quarks and leptons are assigned $R=1$, while those
associated with the Higgs are assigned $R$ charge zero.
Gauginos have $R=+1$.  We want to add an octet and a triplet
field to the model.  A moment's thought indicates that
it is necessary to add at least two triplets to the model
if one is to avoid massless particles.  The problem is that in
the neutralino sector, with only one triplet (taking, for a moment,
the triplet to have $R=0$) there are two positively charged, left-handed
fermions with $R=-1$, while there is only one with $R=+1$;
similar problems arise in the other charge sectors.  This problem
can be solved if we include two triplets in the model, one
with $R=2$, and one with $R=0$.  These will be denoted by
$\hat{T}$ and $\hat{T}^{\prime}$, respectively.  The additional terms
allowed in the superpotential are then
$$W_T = G \hat{T}^a \hat{H}_2 \epsilon \tau^a \hat{H}_1 + B \hat{T}'^a
\hat{T}^a$$

   The scalar  potential generated by this superpotential is:
$$V=\left({g'^2+g^2 \over 32}\right)(H_2^2-H_1^2)^2 +
{G^2\over 4}(H_1^2H_2^2 +
 {T^2H_1^2}+{T^2H_2^2}) + {B^2\over 2}(T^2+T'^2) +
{G B t' H_1 H_2\over 2\sqrt{2}}$$
 $$+ V_{soft}(H_1,H_2)$$
Because of the term linear in $T'$, $T'$ acquires
   a vacuum expectation value:
$$\VEV{T'}= -{G a_1 a_2\over 2 B}\equiv t'.$$  In addition to the
superpotential
we can have an
 R-invariant soft-term in the Lagrangian of the form
$$\L_{soft} = A \tilde{T}'^a \lambda^a$$
where the $\lambda$'s are the fermionic partner of the gauge
bosons, i.e. the gauginos, and $\tilde{T}'$ is the fermionic component of
   the superfield $\hat{T}'$.

The $\rho$ parameter,
        $$ \rho = {M_W^2 \over M_Z^2 \cos^2\theta_W}$$
is given at tree level by$$\rho = {\sum_i v_i^2[I(I+1) -{Y^2\over 4}]
 \over \sum_i v_i^2{Y^2 \over 2} }$$ where I is the weak isospin and Y is
 the hypercharge
 of the scalar multiplet.  The $\rho$ parameter for our case, with a
    triplet and two Higgs doublets, is given by:
$$\rho = {v_1^2 + v_2^2 +
 4 t'^2
 \over v_1^2 + v_2^2} = 1 + {4 t'^2 \over  v_1^2 + v_2^2}$$
Since the $\rho$
   parameter is known to equal 1 to within about 1\%, the vev of the
triplet must be of the order of 12 GeV or less.

The fermion mass matrix, which is our principal concern, divides
into two charged matrices, each $2 \times 2$, and a neutral
$3 \times 3$ matrix.  To avoid phenomenological difficulties
we require that the lightest eigenvalue of each of the charged mass
matrices
is more than half the $Z$ mass.  The lightest neutral should either
have mass greater than about half the $Z$ mass, or should couple
sufficiently weakly to the $Z$ that it does not give too large
a contribution to the $Z$ width.
We have examined various ranges
of parameters, and found that it is possible to satisfy all of these
constraints.  The constraint on the charged masses is easy to
satisfy.  It is more difficult to avoid light neutral particles.
Indeed, study of the mass matrix reveals that the lightest
neutral is never more massive than $\sin(\theta_W)M_Z$;
this bound is saturated for
$A,B\gg M_W$ and $G\gg g,g'$.
For example,
\settabs 6 \columns
\+& $A$  & $B$ & $G$ & $v_1/v_2$ & $m_l$\cr
\+& 9000  & 900 & 1 & 1 & 44\cr
\+& 900  & 900 & 10 & .01 & 44\cr
On the other hand, for a wide range of parameters,
this light state contains nearly equal admixtures of
the two Higgsinos, $\tilde H_1$ and $\tilde H_2$.
Because these fields couple to the $Z$ with opposite
signs, the coupling of this particle to the $Z$ is suppressed.

\bigskip
\chapter{\bf Conclusions}
\bigskip

There are a number of lessons to be drawn from this work.
First, it does not seem so difficult to build models
in which supersymmetry is dynamically broken.  The price
one pays is the introduction of light states beyond
those of the minimal supersymmetric standard model.
In addition, one requires that hypercharge be unified
within a larger group.  Needless to say, it is not clear
whether such an approach will fit neatly into
conventional grand unification or string theory.

We have also commented on some ideas of Hall and Randall
for constructing theories with unbroken R symmetries.
These have the potential to solve some of the other naturalness
problems of supersymmetric theories.  We have noted that
in the context of supergravity, it is difficult to understand
both vanishing cosmological constant and the existence of R symmetries.
Ignoring this question, we have considered various aspects of these
theories,
and have noted that it may be necessary to
add light triplets in the low energy theory to obtain
phenomenologically viable models.  We have seen that models
of this type almost always yield new, relatively light fermions with
interesting properties.

It is perhaps of interest to comment on one other set
of naturalness issues as we close this paper.
Cosmological and astrophysical constraints suggest that
the axion decay constant is in the range $10^{11}-10^{12}$ GeV.
Since this scale is similar to the scale $M_{int}$, it is
natural to ask whether these two scales might be related.
Indeed, this possibility was suggested some time ago
by Kim,\refmark{\kim} and its possible
cosmological significance has been considered by Rajagopal, Turner
and Wilczek.\refmark{\wt}  However, while this coincidence is
tantalizing, it is also problematic.   In the models
considered by Kim, for example, the axion couples to
quarks with masses of order $M_{int}$.  The axion, however,
also couples (with dimensionless couplings) to the `Goldstino'
(the longitudinal component of the gravitino).  As a result,
in this model, there are diagrams at three loop order which
involve only dimensionless couplings and give mass to squarks.
These masses are of order
$$m_{\tilde q}^2 \sim {\alpha_s \over \pi}^2 {\lambda^2 \over 16 \pi^2}
M_{int}^2 \eqn\squarkmass$$
Here $\lambda$ describes the coupling of the axino multiplet
to the goldstino multiplet.  Unless $\lambda$ is extremely small
($\lambda < 10^{-6}$ or so), squarks will obtain unacceptably
large masses.

This problem appears quite general.  In order to link
supersymmetry breaking directly with the axion, the axion
multiplet must couple to the goldstone multiplet.  On the other
hand, in order to have the correct coupling to $F \tilde F$,
the axion must couple to fields carrying color.  But this means
that the hidden sector is not really hidden; while gauge fields
may only couple to the hidden sector through loops, these
couplings are not suppressed by factors of $1 \over M_P$.
There is, of course, no problem in simply introducing the axion
multiplet separately, with no (dimensionless) couplings to the
hidden sector.  However, in this case one needs some other way to
understand
\REF\cincinnati{M. Dine, Talk at the Cincinnati Symposium in honor
of Louis Witten, to appear.}
the coincidence of scales.\refmark{\cincinnati}

\bigskip
{\bf Acknowledgements}

We wish to thank L. Hall and L. Randall for conversations about
their work.  We especially wish to thank Vadim Kaplunovsky
for discussions of problems associated with anomalies, and
for suggesting that various non-supersymmetric results
even in apparently finite calculations could result from
a need for regularization.
This work supported in part by
DOE contract DE-AC02-83ER40107.

\refout
\end